\documentclass[nohyper,12pt,letterpaper]{JHEP3}
\usepackage{graphicx,epsfig,youngtab}

\author{Robert de Mello Koch$^{1,2}$, Badr Awad Elseid Mohammed$^{1}$ and Stephanie Smith$^{1}$\\
$^{1}$ National Institute for Theoretical Physics,\\
Department of Physics and Centre for Theoretical Physics,\\ 
University of the Witwatersrand,\\ 
Wits, 2050, South Africa\\
\qquad\\
$^{2}$Stellenbosch Institute for Advanced Studies,\\
Stellenbosch, South Africa\\
\qquad\\
E-mail: \email{robert@neo.phys.wits.ac.za, Stephanie.Smith@students.wits.ac.za, Badr.Mohammed@students.wits.ac.za}}

\abstract{We compute the one loop anomalous dimensions of restricted Schur polynomials with a classical 
dimension $\Delta\sim O(N)$. The operators that we consider are labeled by Young diagrams with two long 
columns or two long rows. Simple analytic expressions for the action of the dilatation operator are found. 
The projection operators needed to define the restricted Schur polynomials are constructed by translating
the problem into a spin chain language, generalizing earlier results obtained in the $SU(2)$
sector of the theory. The diagonalization of the dilatation operator reduces to solving five term recursion
relations. The recursion relations can be solved exactly in terms of products of symmetric Kravchuk polynomials 
with Hahn polynomials. This proves that the dilatation operator reduces to a decoupled set of harmonic
oscillators and therefore it is integrable, extending a similar conclusion reached for the $SU(2)$
sector of the theory.}

\preprint{WITS-CTP-075}

\title{Nonplanar Integrability: Beyond the $SU(2)$ Sector}


\def \Tr{\mbox{Tr\,}}

\begin{document}

\section{Introduction}

The discovery of integrability in the planar limit of ${\cal N}=4$ super Yang-Mills theory\cite{Minahan:2002ve} has 
lead to tremendous rapid progress (for a recent review see \cite{Beisert:2010jr} and in particular 
\cite{Kristjansen:2010kg,Zoubos:2010kh} which are particularly relevant for our study). 
Non-planar corrections to the planar limit seem to spoil integrability(see for example \cite{Beisert:2003tq,Kristjansen:2008ib}). Does 
this imply that integrability is a property only of the planar limit? In this article we would like to provide evidence 
that this is not the case. We study the large $N$ limit of a set of operators whose bare dimension is of order $N$. For this class of 
operators, the planar approximation does not give an accurate description of the large $N$ limit and one is forced to tackle
the problem of summing an infinite number of non-planar corrections.

There are good reasons to hope that various large $N$ limits are ultimately described by simple physics. In \cite{us} the 
BMN operators\cite{Berenstein:2002jq} in an LLM background\cite{Lin:2004nb} were considered. There is a limit in which the 
resulting dilatation operator commutes with a nontrivial set of conserved charges. In \cite{Koch:2010gp,VinceKate} the spectrum 
of anomalous dimensions of operators AdS/CFT dual\cite{Maldacena:1997re} to giant gravitons\cite{McGreevy:2000cw} was considered. 
The operators considered all belong to the $SU(2)$ sector of the theory. The resulting numerical spectra suggest that the dilatation 
operator reduces to a set of decoupled harmonic oscillators. Motivated by these numerical results, \cite{Carlson:2011hy} studied the 
class of restricted Schur polynomials with two rows/columns. By taking a certain limit, a remarkable simplification takes place. In 
particular, the problem of computing the projectors needed to define the restricted Schur polynomials can be translated into a spin 
chain problem. This allowed an analytic demonstration that the spectrum dilatation operator reduces to that of a set of decoupled harmonic 
oscillators.

{\vskip 0.1cm}

{\sl 
The main goal of this article is to extend the results of \cite{Koch:2010gp,VinceKate,Carlson:2011hy} beyond the $SU(2)$ sector. We find 
that the previous results generalize nicely and we can again give an analytic demonstration that the spectrum of the dilatation operator 
reduces to that of a set of decoupled harmonic oscillators.}

{\vskip 0.1cm}

In the next section we derive an analytic expression for the action of the one loop dilatation operator on restricted Schur polynomials
built using three complex scalars. This is a new result and generalizes the result for the $SU(2)$ sector obtained in \cite{VinceKate}.
In section 3 we describe our construction of the projection operators needed to define the restricted Schur polynomials. We focus on
restricted Schur polynomials labeled by Young diagrams that have two rows/columns. The relevant projectors project from an irreducible 
representation of $S_{n+m+p}$ to an irreducible representation of an $S_n\times S_m\times S_p$ subgroup. For two rows/columns a
given irreducible $S_n\times S_m\times S_p$ representations is subduced at most once from a given $S_{n+m+p}$ irreducible representation.
As discussed in \cite{VinceKate} this simplifies the problem of computing the projectors significantly.
Our construction trades the problem of constructing the projector for the eigenproblem of certain $S_m\times S_p$ Casimirs.
This eigenproblem is then solved by translating it into a spin chain language, significantly generalizing the construction 
of \cite{Carlson:2011hy}. In section 4 we use our construction of the projection operators to obtain explicit formulas for the action 
of the dilatation operator. This evaluation is a little more than an application of the simple theory of addition of angular momentum in
ordinary non-relativistic quantum mechanics. The eigenproblem of the dilatation operator is solved in section 5 and we discuss our results 
in section 6. In Appendix A we give a detailed construction of a projection operator using the new spin chain method. In Appendix B we 
summarize the representation theory needed to understand our construction. Appendix C provides a numerical study of the the one loop 
dilatation operator on restricted Schur polynomials built using three complex scalars.

\section{Action of the Dilatation Operator}

{\sl In this section we will study the action of the one loop dilatation operator on restricted Schur polynomials
built using three complex adjoint scalars. The main result of this section, which generalizes results known for the
$SU(2)$ sector\cite{VinceKate}, is the simple formula (\ref{dilat}) for the action of the dilatation operator.}

{\vskip 0.2cm}

Our operators are built using the six scalar fields $\phi^i$, which take values in the adjoint of $u(N)$ in ${\cal N}=4$ 
super Yang Mills theory. Assemble these scalars into the three complex combinations
$$
  Z=\phi_1+i\phi_2 ,\qquad Y=\phi_3+i\phi_4 ,\qquad X=\phi_5+i\phi_6 \, .
$$
The operators we consider are built using $O(N)$ of these complex scalar fields. These operators have a large ${\cal R}$-charge and consequently, 
non-planar contributions to the correlation functions of these operators are not suppressed at large $N$\cite{Balasubramanian:2001nh}. 
The computation of the anomalous dimensions of these operators is then a problem of considerable complexity. This problem has been 
effectively handled by new methods which employ group representation 
theory\cite{Corley:2001zk,SS,Balasubramanian:2004nb,de Mello Koch:2007uu,de Mello Koch:2007uv,Kimura:2007wy,Bekker:2007ea,Brown:2007xh,Bhattacharyya:2008rb,Brown:2008rr,Kimura:2008wy,Kimura:2009wy,Ramgoolam:2008yr}
allowing one to sum all diagrams (planar and non-planar) contributing. Indeed, the two point function of restricted Schur 
polynomials\cite{Balasubramanian:2004nb,de Mello Koch:2007uu,de Mello Koch:2007uv,Bekker:2007ea} can be evaluated exactly in the free 
field theory limit\cite{Bhattacharyya:2008rb}. The restricted Schur polynomials provide a basis for the local 
operators\cite{Bhattacharyya:2008rc} which diagonalize the free two point function and which have highly constrained mixing at the 
quantum level\cite{de Mello Koch:2007uv,Bekker:2007ea,Koch:2010gp,VinceKate,Carlson:2011hy}.
For the applications that we have in mind, this basis is clearly far superior to the trace basis. Mixing between operators 
in the trace basis with this large ${\cal R}$-charge is completely unconstrained even at the level of the free theory.

The restricted Schur polynomials are
$$
\chi_{R,(r,s,t)}(Z^{\otimes \, n},Y^{\otimes \, m},X^{\otimes\, p}) =
{1\over n!m! p!}\sum_{\sigma\in S_{n+m+p}}{\rm Tr}_{(r,s,t)}(\Gamma_R(\sigma))
X^{i_1}_{i_{\sigma(1)}}\cdots X^{i_p}_{i_{\sigma(p)}}\times
$$
$$
\times Y^{i_{p+1}}_{i_{\sigma(p+1)}}\cdots Y^{i_{p+m}}_{i_{\sigma(p+m)}}
Z^{i_{p+m+1}}_{i_{\sigma(p+m+1)}}\cdots Z^{i_{n+m+p}}_{i_{\sigma(n+m+p)}}\, .
$$
We use $n$ to denote the number of $Z$s, $m$ to denote the number of $Y$s and $p$ to denote the number of $X$s.
$R$ is a Young diagram with $n+m+p$ boxes or equivalently an irreducible representation of $S_{n+m+p}$. $r$ is a 
Young diagram with $n$ boxes or equivalently an irreducible representation of $S_n$, $s$ is a Young diagram with 
$m$ boxes or equivalently an irreducible representation of $S_m$ and $t$ is a Young diagram with $p$ boxes or 
equivalently an irreducible representation of $S_p$. The $S_n$ subgroup acts on $m+p+1,m+p+2,...,m+p+n$ and therefore 
permutes indices belonging to the $Z$s. The $S_m$ subgroup acts on $p+1,p+2,...,p+m$ and hence permutes indices belonging to 
the $Y$s. The $S_p$ subgroup acts on $1,2,...,p$ and hence permutes indices belonging to the $X$s.
Taken together $(r,s,t)$ specify an irreducible representation of $S_n\times S_m\times S_p$. ${\rm Tr}_{(r,s,t)}$
is an instruction to trace over the subspace carrying the irreducible representation\footnote{In general, because $(r,s,t)$
can be subduced more than once, we should include a multiplicity index. We will not write or need this index in this article.
We will, in the next section, restrict our attention to restricted Schur polynomials that are labeled by Young diagrams with
two rows or columns. A huge simplification that results is that all possible representations $(r,s,t)$ are subduced exactly once.}
$(r,s,t)$ of $S_n\times S_m\times S_p$ inside the carrier space for irreducible representation $R$ of $S_{n+m+p}$.
This trace is easily realized by including a projector $P_{R\to (r,s,t)}$ (from the carrier space of $R$ to the carrier 
space of $(r,s,t)$) and tracing over all of $R$, i.e. ${\rm Tr}_{(r,s,t)}(\Gamma_R(\sigma))={\rm Tr}(P_{R\to(r,s,t)}\Gamma_R(\sigma))$.

The one loop dilatation operator, when acting on operators composed from the three complex scalars $X,Y,Z$, 
is\cite{Kristjansen:2002bb,Constable:2002hw,Constable:2002vq,Beisert:2002tn,Minahan:2002ve,Beisert:2002ff}
$$
D = - g_{\rm YM}^2 {\rm Tr}\,\big[ Y,Z\big]\big[ \partial_Y ,\partial_Z\big]
    - g_{\rm YM}^2 {\rm Tr}\,\big[ X,Z\big]\big[ \partial_X ,\partial_Z\big]
    - g_{\rm YM}^2 {\rm Tr}\,\big[ Y,X\big]\big[ \partial_Y ,\partial_X\big].
$$
The action of the dilatation operator on the restricted Schur polynomials belonging to the $SU(2)$ sector has been worked out in 
\cite{Koch:2010gp,VinceKate}. In what follows, we will work with operators normalized to give a unit two point function. The two 
point functions for restricted Schur polynomials has been computed in \cite{Bhattacharyya:2008rb}
$$
\langle\chi_{R,(r,s,t)}(Z,Y)\chi_{T,(u,v,w)}(Z,Y)^\dagger\rangle =
\delta_{R,(r,s,t)\,T,(u,v,w)}f_R {{\rm hooks}_R\over {\rm hooks}_r \, {\rm hooks}_s\, {\rm hooks}_t}\, .
$$
In this expression $f_R$ is the product of the factors\footnote{The term weights is also frequently used. The factor/weight of a box 
in the $i^{\rm th}$ row and $j^{\rm th}$ column is $N+j-i$.} in Young diagram $R$ and ${\rm hooks}_R$ is the product of the hook
lengths of Young diagram $R$. Thus, the normalized operators $O_{R,(r,s,t)}(Z,Y)$ can be obtained from
$$
\chi_{R,(r,s,t)}(Z,Y,X)=\sqrt{f_R \, {\rm hooks}_R\over {\rm hooks}_r\, {\rm hooks}_s\, {\rm hooks}_t}O_{R,(r,s,t)}(Z,Y,X)\, .
$$
The computation of the dilatation operator is a simple extension of the analysis presented in \cite{VinceKate} so that we will only 
quote the final result. In terms of the normalized operators
\begin{equation}
DO_{R,(r,s,t)}(Z,Y,X)=\sum_{T,(u,v,w)} N_{R,(r,s,t);T,(u,v,w)}O_{T,(u,v,w)}(Z,Y,X)
\label{dilat}
\end{equation}
{\small
\begin{eqnarray}
\nonumber
N_{R,(r,s,t);T,(u,v,w)}= - \sum_{R'}{c_{RR'}g_{YM}^2 d_T \over d_{R'} d_u d_v d_w(n+m+p)}
\sqrt{f_T \,{\rm hooks}_T\,{\rm hooks}_r\,{\rm hooks}_s {\rm hooks}_t\over f_R\,{\rm hooks}_R\,{\rm hooks}_u\,{\rm hooks}_v\,{\rm hooks}_w}\times
\\
\nonumber
\times\left[nm\Tr\Big(\Big[ \Gamma_R((p+m+1,p+1)),P_{R\to (r,s,t)}\Big]I_{R'\, T'}\Big[\Gamma_T((p+m+1,p+1)),P_{T\to (u,v,w)}\Big]I_{T'\, R'}\Big) 
\right. \\
\nonumber
+np\Tr\Big(\Big[ \Gamma_R((1,p+m+1)),P_{R\to (r,s,t)}\Big]I_{R'\, T'}\Big[\Gamma_T((1,p+m+1)),P_{T\to (u,v,w)}\Big]I_{T'\, R'}\Big)
\\
\nonumber
+\left.mp\Tr\Big(\Big[ \Gamma_R((1,p+1)),P_{R\to (r,s,t)}\Big]I_{R'\, T'}\Big[\Gamma_T((1,p+1)),P_{T\to (u,v,w)}\Big]I_{T'\, R'}\Big) \right]\, .
\end{eqnarray}
}
$c_{RR'}$ is the factor of the corner box removed from Young diagram $R$ to obtain diagram $R'$, and similarly $T'$ is a Young
diagram obtained from $T$ by removing a box. This factor arises after using the reduction rule of \cite{de Mello Koch:2004ws,de Mello Koch:2007uu}.
The intertwiner $I_{AB}$ is a map from the carrier space of irreducible representation $A$ to the carrier space of irreducibe representation $B$. 
Consequently, by Schur's Lemma, $A$ and $B$ must be Young diagrams of the same shape. The intertwiner operators relevant for our study have been 
discussed in detail in \cite{VinceKate}.

\section{Projection Operators}

{\sl The goal of this section is to construct the projection operators needed to define the restricted Schur polynomials
we study in this article. This construction clearly defines the class of operators being considered. The approximations
being employed in this construction are carefully considered.}

{\vskip 0.2cm}

The class of operators $\chi_{R,(r,s,t)}(Z,Y,X)$ we will study in this article are labeled by
Young diagrams that each have 2 rows or columns. We further take $n$ to be order $N$ and 
$m,p$ to be $\alpha N$ with $\alpha\ll 1$. Thus, there are a lot more $Z$ fields than there are
$Y$s or $X$s. The mixing of these operators with restricted Schur polynomials that have $n\ne 2$ 
rows or columns (or of even more general shape) is suppressed at least by a factor of order ${1\over\sqrt{N}}$\footnote{Here we are 
talking about mixing at the quantum level. There is no mixing in the free theory\cite{Bhattacharyya:2008rb}.}.
Thus, at large $N$ the 2 row or column restricted Schur polynomials do not mix with other operators, which is a huge simplification.
This is the analog of the statement that for operators with a dimension of $O(1)$, different trace structures do not mix at large $N$.
The fact that the two column restricted Schur polynomials are a decoupled sector at large $N$ is expected: these operators correspond
to a well defined stable semi-classical object in spacetime (the two giant graviton system).

Note that as a consequence of the fact that there are a lot more $Z$s than $Y$s and $X$s, contributions to the dilatation
operators coming from interactions between $Z$s and $Y$s or between $Z$s and $X$s will over power the contribution coming from
interactions between $X$s and $Y$s. Consequently we can simplify the action of the dilatation operator to
{\small
\begin{eqnarray}
\label{evlaute}
&&N_{R,(r,s,t);T,(u,v,w)}= - \sum_{R'}{c_{RR'}g_{YM}^2 d_T n\over d_{R'} d_u d_v d_w(n+m+p)}
\sqrt{f_T \,{\rm hooks}_T\,{\rm hooks}_r\,{\rm hooks}_s {\rm hooks}_t\over f_R\,{\rm hooks}_R\,{\rm hooks}_u\,{\rm hooks}_v\,{\rm hooks}_w}\times
\\
\nonumber
&&\times\left[m\Tr\Big(\Big[ \Gamma_R((p+m+1,p+1)),P_{R\to (r,s,t)}\Big]I_{R'\, T'}\Big[\Gamma_T((p+m+1,p+1)),P_{T\to (u,v,w)}\Big]I_{T'\, R'}\Big) \right.
\\
\nonumber
&&+\left. p\Tr\Big(\Big[ \Gamma_R((1,p+m+1)),P_{R\to (r,s,t)}\Big]I_{R'\, T'}\Big[\Gamma_T((1,p+m+1)),P_{T\to (u,v,w)}\Big]I_{T'\, R'}\Big) \right]\, .
\end{eqnarray}
}
We will obtain an analytic expression for the above operator in this article.

\subsection{Two Rows}

We will make use of Young's orthogonal representation for the symmetric group.
This representation is most easily defined by considering the action of adjacent
permutations (permutations of the form $(i,i+1)$) on the Young-Yamonouchi states.
The permutation $(i,i+1)$ when acting on any given Young-Yamonouchi state will
produce a linear combination of the original state and the state obtained by 
swapping the positions of $i$ and $i+1$ in the Young-Yamonouchi symbol. The
precise rule is most easily written in terms of the axial distance between $i$
and $i+1$. If $i$ appears in row $r_i$ and column $c_i$ of the Young-Yamonouchi symbol
and $i+1$ appears in row $r_{i+1}$ and column $c_{i+1}$ of the Young-Yamonouchi symbol,
then the axial distance between $i$ and $i+1$ is
$$
  d_{i,i+1}=c_i-r_i - (c_{i+1}-r_{i+1})\, .
$$
In terms of this axial distance, the action of $(i,i+1)$ is 
$$
  (i,i+1)\left|{\rm state}\right\rangle = {1\over d_{i,i+1}}\left|{\rm state}\right\rangle
       +\sqrt{1-{1\over d_{i,i+1}^2}}\left|{\rm swapped\,\, state}\right\rangle
$$
where the Young-Yamonouchi symbol of $\left|{\rm swapped\,\, state}\right\rangle$ state is obtained
from the Young-Yamonouchi symbol of $\left|{\rm state}\right\rangle$ by swapping the positions of
$i$ and $i+1$. See \cite{hammermesh} for more details.

The reason why we use Young's orthogonal representation is that it simplifies dramatically for
the operators we are interested in. To construct the projectors $P_{R\to (r,s,t)}$ we will imagine
that we start by removing $m+p$ boxes from $R$ to produce $r$. We label the boxes in the order that they
are removed. Of course, after each box is removed we are left with a valid Young diagram; this is
a nontrivial constraint on the allowed numberings. Thus, after labeling these boxes we have a total 
of $2^{m+p}$ partially labeled Young diagrams, each corresponding to a subspace $r$ of the subgroup 
$S_n\times (S_1)^{m+p}$ of the original $S_{n+m+p}$ group. We now need to take linear combinations 
of these subspaces in such a way that we obtain the correct irreducible representation $(s,t)$ of 
the $S_m\times S_p$ subgroup that acts on the labeled boxes. For the class of operators that we 
consider, the number of boxes that we remove ($=m+p$) is much less that the number of boxes in $R$ 
($=m+n+p\approx n$). In the figure below we show $R$ and the boxes that must be removed from $R$ to
obtain $r$. It is clear that the axial distance $d_{i,i+1}$ is 1 if the boxes are in the same row
so that
$$
  (i,i+1)\left|{\rm state}\right\rangle = \left|{\rm state}\right\rangle
       \qquad {\rm for \,\, boxes \,\, in \,\, the \,\, same \,\, row}\, .
$$
It is also clear that $d_{i,i+1}$ is $O(N)$ for boxes in different rows. At large $N$ we can
simply set $(d_{i,i+1})^{-1}=0$ so that
$$
  (i,i+1)\left|{\rm state}\right\rangle = \left|{\rm swapped\,\, state}\right\rangle
       \qquad {\rm for \,\, boxes \,\, in \,\, different \,\, rows}\, .
$$
\begin{figure}[h]
          \centering
          {\epsfig{file=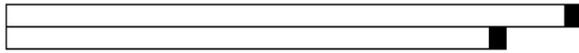,width=8.0cm,height=1cm}}
          \caption{Shown above is the Young diagram $R$. The boxes that are to be removed from $R$
                   to obtain $r$ are colored black.}
\end{figure}
The representation that we have obtained is very similar to a representation which has already been 
studied in the mathematics literature \cite{GelPair}. Motivated by this background define a map from 
a labeled Young diagram to a monomial. Our Young diagram has $m+p$ boxes labeled and the labels are 
distributed between the upper and lower rows. Ignore the boxes that appear in the lower row. For boxes
labeled $i$ in the upper row include a factor of $x_i$ in the monomial if $1\le i\le p$ and a factor
of $y_i$ if $p+1\le i\le p+m$. If none of the boxes in the first row are labeled, the Young diagram maps 
to 1. Thus, for example, when $m=2$ and $p=2$
{\small
$$
{\small \young({\,}{\,}{\,}{\,}{\,}{\,}{\,}{\,}{\,}{3},{\,}{\,}{4}{2}{1})}\leftrightarrow y_3
\qquad
{\small \young({\,}{\,}{\,}{\,}{\,}{\,}{\,}{3}{2}{1},{\,}{\,}{\,}{\,}{4})}\leftrightarrow
x_1 x_2 y_3
$$
}
The symmetric group acts by permuting the labels on the factors in the monomial. Thus, for example,
$ (12)x_1 y_3 = x_2 y_3 $. This defines a reducible representation of the group $S_m\times S_p$. 
It is clear that the operators\footnote{It may be helpful (and it is accurate) for the reader to
associate the $x_i,y_j$ of these operators with the $X^i_{\sigma(i)},Y^j_{\sigma(j)}$ appearing
in the definition of the restricted Schur polynomials.}
\begin{equation}
d_1=\sum_{i=1}^p {\partial\over \partial x_i} \qquad d_2=\sum_{i=p+1}^{p+m} {\partial\over \partial y_i}
\end{equation}
commute with the action of the $S_m\times S_p$ subgroup. These operators generalize closely related operators introduced 
by Dunkl in his study of intertwining functions \cite{Dunkl}. They act on the monomials by producing the sum 
of terms that can be produced by dropping one $x$ factor for $d_1$ or one $y$ factor for $d_2$ at a time. For example
$$
d_1 (x_1 x_2 y_3) = x_2 y_3 + x_1 y_3 ,\qquad d_2 (x_1 x_2 y_3) = x_1 x_2 \, .
$$
The adjoint\footnote{Consult Appendix \ref{rep} for details on the inner product on the space of monomials.} produces the
sum of monomials that can be obtained by appending a factor, without repeating any of the factors (this is written for $m=2=p$
impurities but the generalization to any $m$ is obvious)
$$
  d_1^\dagger (y_3)= x_1 y_3 + x_2 y_3,\qquad  d_1^\dagger (x_1 y_3)= x_1 x_2 y_3,\qquad  d_2^\dagger (x_1 y_3)= x_1 y_3 y_4\, .
$$
The fact that $d_1$ and $d_2$ commute with all elements of $S_m\times S_p$, implies that $d_1^\dagger$ and $d_2^\dagger$ will too. 
Thus, $d_1^\dagger d_1$ and $d_2^\dagger d_2$ will also commute with all the elements of the $S_m\times S_p$ subgroup and consequently 
their eigenspaces will furnish representations of the subgroup. These eigenspaces are irreducible representations - consult 
\cite{GelPair} for useful details and results. This last fact implies that the problem of computing the projectors needed to define 
the restricted Schur polynomials can be replaced by the problem of constructing projectors onto the eigenspaces of $d_1^\dagger d_1$
and $d_2^\dagger d_2$. This amounts to solving for the eigenvectors and eigenvalues of $d_1^\dagger d_1$ and $d_2^\dagger d_2$. This 
problem is most easily solved by mapping the labeled Young diagrams into states of a spin chain. The spin at site $i$ can be in state 
spin up ($+{1\over 2}$) or state spin down ($-{1\over 2}$). The spin chain has $m+p$ sites and the box labeled $i$ tells us the state 
of site $i$. If box $i$ appears in the first row, site $i$ is in state $+{1\over 2}$; if it appears in the second row site $i$ is in 
state $-{1\over 2}$. For example,
{\small
$$
\young({\,}{\,}{\,}{\,}{\,}{\,}{\,}{\,}{\,}{5}{2}{1},{\,}{\,}{\,}{\,}{6}{4}{3})\leftrightarrow 
\left|{1\over 2},{1\over 2},-{1\over 2},-{1\over 2},{1\over 2},-{1\over 2}\right\rangle
$$
}
Both $d_1^\dagger d_1$ and $d_2^\dagger d_2$ have a very simple action on this spin chain: Introduce the states
$$
\left| {1\over 2}\right\rangle =\left[ \begin {array}{c} 1\\\noalign{\medskip}0\end {array} \right] \qquad
\left|-{1\over 2}\right\rangle =\left[ \begin {array}{c} 0\\\noalign{\medskip}1\end {array} \right]
$$
for the possible states of each site and the operators
$$
  \sigma^+                      = \left[ \begin {array}{cc} 0&1\\\noalign{\medskip}0&0\end {array} \right] \qquad
  \sigma^- = (\sigma^+)^\dagger = \left[ \begin {array}{cc} 0&0\\\noalign{\medskip}1&0\end {array} \right]
$$
which act on these states
$$
  \sigma^+ \left|-{1\over 2}\right\rangle = \left|{1\over 2}\right\rangle, \qquad
  \sigma^+ \left|{1\over 2}\right\rangle = 0, \qquad 
  \sigma^- \left|{1\over 2}\right\rangle = \left|-{1\over 2}\right\rangle, \qquad  
  \sigma^- \left|-{1\over 2}\right\rangle = 0\, .
$$
We can write any of the states of the spin chain as a tensor product of the states 
$|{1\over 2}\rangle$ and $|-{1\over 2}\rangle$. For example 
$$
  \left|-{1\over 2},-{1\over 2},{1\over 2},-{1\over 2},{1\over 2},{1\over 2}\right\rangle = 
  \left|-{1\over 2}\right\rangle\otimes\left|-{1\over 2}\right\rangle\otimes\left|{1\over 2}\right\rangle\otimes
  \left|-{1\over 2}\right\rangle\otimes\left| {1\over 2}\right\rangle\otimes\left|{1\over 2}\right\rangle
$$
for a system with 6 lattice sites. Label the sites starting from the left, as site 1, then site 2 and so on till we get
to the last site, which is site 6. The operator $\sigma^-$ acting at the third site (for example) is
$$
  \sigma^-_3 = 1\otimes 1\otimes \sigma^- \otimes 1\otimes 1\otimes 1\, .
$$

We can then write
\begin{eqnarray}
  d_1^\dagger d_1 && = \sum_{\alpha =1}^p \sum_{\beta=1}^p \sigma^+_\alpha \sigma^-_\beta \, ,\\
  d_2^\dagger d_2 && = \sum_{\alpha =p+1}^{p+m} \sum_{\beta=p+1}^{p+m} \sigma^+_\alpha \sigma^-_\beta \, 
\end{eqnarray}
This is a long ranged spin chain. In terms of the Pauli matrices
$$
  \sigma^1=\left[ \begin {array}{cc} 0&1\\\noalign{\medskip}1&0\end {array} \right] \qquad
  \sigma^2=\left[ \begin {array}{cc} 0&-i\\\noalign{\medskip}i&0\end {array} \right]\qquad
  \sigma^3=\left[ \begin {array}{cc} 1&0\\\noalign{\medskip}0&-1\end {array} \right]
$$
we define the following ``total spins'' of the system
$$
J^{1}=\sum_{\alpha =1}^p\frac{1}{2}\sigma_{\alpha}^{1}\, ,\,\,\,
J^{2}=\sum_{\alpha =1}^p\frac{1}{2}\sigma_{\alpha}^{2}\,,\,\,\,
J^{3}=\sum_{\alpha =1}^p\frac{1}{2}\sigma_{\alpha}^{3}\, ,
$$
$$
\textbf{J}^{2}=J^{1}J^{1}+J^{2}J^{2}+J^{3}J^{3}\, ,
$$
and
$$
K^{1}=\sum_{\alpha =p+1}^{p+m}\frac{1}{2}\sigma_{\alpha}^{1}\,,\,\,\,
K^{2}=\sum_{\alpha =p+1}^{p+m}\frac{1}{2}\sigma_{\alpha}^{2}\,,\,\,\,
K^{3}=\sum_{\alpha =p+1}^{p+m}\frac{1}{2}\sigma_{\alpha}^{3}\, ,
$$
$$
\textbf{K}^{2}=K^{1}K^{1}+K^{2}K^{2}+K^{3}K^{3}\, .
$$
We use capital letters for operators and little letters for
eigenvalues. In terms of these total spins we have
$$
  d_1^\dagger d_1  = \textbf{J}^{2}-J^{3}(J^{3}+1)\,,\qquad d_2^\dagger d_2  = \textbf{K}^{2}-K^{3}(K^{3}+1)\,.
$$
Thus, eigenspaces of $d_1^\dagger d_1$ can be labeled by the
eigenvalues of $\textbf{J}^2$ and eigenvalues of $J^3$, and 
the eigenspaces of $d_2^\dagger d_2$ can be labeled by the
eigenvalues of $\textbf{K}^2$ and eigenvalues of $K^3$.
Consequently, the labels $R,(r,s,t)$ of the restricted Schur polynomial can be
traded for these eigenvalues. Indeed, consider the restricted Schur polynomial 
$\chi_{R,(r,s,t)}(Z,Y,X)$. The $\textbf{K}^2 = k(k+1)$ quantum number tells you the shape of 
the Young diagram $s$ that organizes the impurities: if there are $N_1$ boxes 
in the first row of $s$ and $N_2$ boxes in the second, then $2k=N_1-N_2$. 
The $\textbf{J}^2 = j(j+1)$ quantum tells you the shape of the Young diagram 
$t$ that organizes the impurities: if there are $N_1$ boxes in the first row of 
$t$ and $N_2$ boxes in the second, then $2j=N_1-N_2$. The $J^3+K^3$ eigenvalue
of the state is always a good quantum number, both in the basis we start in 
where each spin has a sharp angular momentum or in the basis where the states 
have two sharp ``total angular momenta''. The $j^3+k^3$ quantum number tells you 
how many boxes must be removed from each row of $R$ to obtain $r$. Denote 
the number of boxes to be removed from the first row by $n_1$ and the number
of boxes to be removed from the second row by $n_2$. We have $2j^3+2k_3=n_1-n_2$.
This gives a complete construction of the projection operators we need.

To get some insight into how the construction works, lets count the states which appear
for the example $m=p=4$. There are three possible Young diagram shapes which appear
$$
  \yng(4)\qquad \yng(3,1)\qquad \yng(2,2)\, .
$$
These correspond to a spins of $2,1,0$ respectively. As irreducible representations of $S_4$ 
they have a dimension of 1, 3 and 2 respectively. Coupling four spins we have
$$
  {\bf {1\over 2}}\otimes {\bf {1\over 2}}\otimes {\bf {1\over 2}}\otimes {\bf {1\over 2}} 
   = {\bf 2}\oplus 3{\bf 1}\oplus 2{\bf 0}\, .
$$
These results illustrate that each state of a definite spin labels an irreducible
representation of the symmetric group and further that for our 8 spins we find the following
organization of states
$$
\begin{array}{cccc}
S_m\times S_p{\rm \,\, irrep\qquad} & \textbf{K}{\rm \,\, irrep\qquad} & \textbf{J}{\rm \,\, irrep\qquad} &{\rm dimension}\\
(\yng(4),\yng(4))                   & k=2                          & j=2                          &  25\\
(\yng(4),\yng(3,1))                 & k=2                          & j=1                          &  45\\
(\yng(4),\yng(2,2))                 & k=2                          & j=0                          &  10\\
(\yng(3,1),\yng(4))                 & k=1                          & j=2                          &  45\\
(\yng(3,1),\yng(3,1))               & k=1                          & j=1                          &  81\\
(\yng(3,1),\yng(2,2))               & k=1                          & j=0                          &  18\\
(\yng(2,2),\yng(4))                 & k=0                          & j=2                          &  10\\
(\yng(2,2),\yng(3,1))               & k=0                          & j=1                          &  18\\
(\yng(2,2),\yng(2,2))               & k=0                          & j=0                          &  4
\end{array}
$$
The last column is obtained by taking a product of the dimension of the $S_m\times S_p$ irreducible representation by the dimension 
$(2k+1)(2j+1)$ of the associated spin multiplets. Summing the entries in the last column we obtain 256 which is indeed the number of 
states in the spin chain. For a detailed example of how the construction works see Appendix \ref{projector}.

{\vskip 0.2cm}

\noindent
{\bf Summary of the Approximations made:}

\begin{itemize}

\item{} We have neglected mixing with restricted Schur polynomials that have $n\ne 2$ rows. These mixing terms are at most
        $O({1\over\sqrt{N}})$ so that this approximation is accurate at large $N$.

\item{} The terms arising from an interaction between the $X$s and $Y$s have been neglected. Since there are a lot more
        $Z$s than $X$s and $Y$s the one loop dilatation operator will be dominated by terms arising from an interaction
        between $Z$s and $X$s and between $Z$s and $Y$s.

\item{} In simplifying Young's orthogonal representation for the symmetric group we have replaced certain factors
        $(d_{i,i+1})^{-1}=O(N^{-1})$ by $(d_{i,i+1})^{-1}=0$. This is valid at large $N$. The fact that $d_{i,i+1}=O(N)$ is
        a consequence of the fact that we have Young diagrams with two rows, that we consider an operator whose bare
        dimension grows parametrically with $N$ and that there are a lot more $Z$s than $X$s and $Y$s. Thus boxes in
        different rows, corresponding to $X$s and $Y$s, are always separated by a large axial distance at large $N$. 

\end{itemize}

\subsection{Two Columns}

To treat the case of two columns, we need to account for the fact that Young's orthogonal representation simplifies to
$$
  (i,i+1)\left|{\rm state}\right\rangle = -\left|{\rm state}\right\rangle
       \qquad {\rm for \,\, boxes \,\, in \,\, the \,\, same \,\, row}\, ,
$$
$$
  (i,i+1)\left|{\rm state}\right\rangle = \left|{\rm swapped\,\, state}\right\rangle
       \qquad {\rm for \,\, boxes \,\, in \,\, different \,\, rows}\, .
$$
Note the minus sign on the first line above. We can account for this sign, generalizing \cite{Carlson:2011hy}, by employing
a description that uses Grassmann variables. To describe the first $p$ boxes, introduce the $2p$ variables $x^+_i,x^-_i$,
where $i=1,2,...,p$. To describe the next $m$ boxes, introduce the $2m$ variables $y^-_j,y^+_j$, where $j=p+1,p+2,...,p+m$.
Each labeled Young diagram continues to have an expression in terms of a monomial. Boxes in the right most column have a
superscript $+$; boxes in the left most column have a superscript $-$. Each monomial is ordered with (i) $x$s to the left
of $y$s and (ii) within each type ($x$ or $y$) of variable, variables with a $-$ superscript to the left of variables with 
a $+$ superscript. Finally within a given type and a given superscript the variables are ordered so that the subscripts
increase from left to right. Thus, for example, when $m=3=p$ we have
$$
\young({\,}{\,},{\,}{5},{\,}{4},{\,}{1},{\,},{6},{3},{2})\leftrightarrow x_1^- x_2^+ x_3^+ y_4^- y_5^- y_6^+\, .
$$
If we now allow $S_m\times S_p$ to act on the monomials by acting on the subscripts of each variable without changing the 
order of the variables, we recover the correct action on the labeled Young diagrams. 

It is a simple matter to show that
$$
  d_1=\sum_{i=1}^p x_i^+{\partial \over\partial x^-_i},\qquad d_2=\sum_{i=p+1}^{p+m} y_i^+{\partial \over\partial y^-_i},
$$
both commute with the symmetric group. It is again simple to show that\footnote{Assuming we only consider monomials that are 
ordered as we described above, the inner product of two identical monomials is 1 and of two different monomials is 0.}
$$
  d_1^\dagger =\sum_{i=1}^p x_i^-{\partial \over\partial x^+_i}, \qquad
  d_2^\dagger =\sum_{i=p+1}^{p+m} y_i^-{\partial \over\partial y^+_i}.
$$
We can again define two $S_m\times S_p$ Casimirs as $d_1^\dagger d_1$ and $d_2^\dagger d_2$. In terms
of the spin variables
$$ 
  \tilde{\sigma}_n^i = (\sigma_n^3)^n \sigma_n^i(\sigma_n^3)^n
$$
we have
$$
  d_1^\dagger d_1  = \tilde{\textbf{J}}^{2}-\tilde{J}^{3}(\tilde{J}^{3}+1)\,,\qquad 
  d_2^\dagger d_2  = \tilde{\textbf{K}}^{2}-\tilde{K}^{3}(\tilde{K}^{3}+1)\,.
$$
Thus, the eigenspaces of $d_1^\dagger d_1$ can be labeled by the eigenvalues of $\tilde{\textbf{J}}^2$ and eigenvalues of $\tilde{J}^3$, and 
the eigenspaces of $d_2^\dagger d_2$ can be labeled by the eigenvalues of $\tilde{\textbf{K}}^2$ and eigenvalues of $\tilde{K}^3$.
Consequently, the labels $R,(r,s,t)$ of the restricted Schur polynomial can again be traded for these eigenvalues. The remaining discussion is
now identical to that of two rows and is thus not repeated.

\section{Evaluation of the Dilatation Operator}

{\sl In this section we will argue that all of the factors in the dilatation operator have a natural interpretation as operators
     acting on the spin chain. This allows us to explicitly evaluate the action of the dilatation operator. Our final formula for
     the dilatation operator is given as the last formula in this section.}

{\vskip 0.2cm}

The bulk of the work involved in evaluating the dilatation operator comes from evaluating the traces
$$
 \Tr\Big(\Big[ \Gamma_R((p+m+1,p+1)),P_{R\to (r,s,t)}\Big]I_{R'\, T'}\Big[\Gamma_T((p+m+1,p+1)),P_{T\to (u,v,w)}\Big]I_{T'\, R'}\Big)\, ,
$$
and
$$
\Tr\Big(\Big[ \Gamma_R((1,p+m+1)),P_{R\to (r,s,t)}\Big]I_{R'\, T'}\Big[\Gamma_T((1,p+m+1)),P_{T\to (u,v,w)}\Big]I_{T'\, R'}\Big) \, .
$$
When we evaluate the second trace above, the intertwiners can be taken to act on the first site of the spin chain. This term corresponds
to an interaction between a $Z$ and $X$ field. The first $p$ sites of the spin chain correspond to $X$ fields so that the intertwiner
could have acted on any of the first $p$ sites of the chain. When we evaluate the first trace above, the intertwiners can be taken to act 
on the $(p+1)$th site of the spin chain. This term corresponds to an interaction between a $Z$ and $Y$ field. The last $m$ sites of the 
spin chain correspond to $Y$ fields so that the intertwiner could have acted on any of the last $m$ sites of the chain. Consider an
intertwiner which acts on the first site of the chain. If the box from row $i$ is dropped from $R$ and the box from row $j$ is dropped 
from $T$, the intertwiner becomes
$$
  I_{R'T'}=E_{ij}\otimes{\bf 1}\otimes \cdots\otimes {\bf 1}\, ,\qquad   I_{T'R'}=E_{ji}\otimes{\bf 1}\otimes \cdots\otimes {\bf 1}\, ,
$$
where $E_{ij}$ is a $2\times 2$ matrix of zeroes except for a 1 in row $i$ and column $j$. We will use a simpler notation according to
which we suppress all factors of the $2\times 2$ identity matrix and indicate which site a matrix acts on by a superscript. Thus, for 
example
$$
  I_{R'T'}=E_{ij}^{(1)},\qquad I_{T'R'}=E_{ji}^{(1)}\, .
$$
Next, consider $\Gamma_R((p+m+1,p+1))$ which acts on a slot occupied by a $Z$ and a slot occupied by a $Y$ and $\Gamma_R((1,p+m+1))$ which acts 
on a slot occupied by a $Z$ and a slot occupied by an $X$. To allow an action on the $Z$ slot, enlarge the spin chain by one extra site (the 
$Z$ site). The projectors and intertwiners all have a trivial action on this $(m+p+1)$th site. $\Gamma_R((p+m+1,p+1))$ will swap the spin in
the $(m+p+1)$th site with the spin in site $p+1$. Thus, we have
$$
  I_{R'T'}\Gamma_R((p+m+1,p+1))=\sum_{k=1}^2 E_{ij}^{(p+1)}E_{kk}^{(m+p+1)}\Gamma_R((p+m+1,p+1))=\sum_{k=1}^2 E_{ik}^{(p+1)}E_{kj}^{(m+p+1)},
$$
$$
  \Gamma_R((p+m+1,p+1))I_{R'T'}=\sum_{k=1}^2 E_{kj}^{(p+1)}E_{ik}^{(m+p+1)},
$$
$$
  \Gamma_R((p+m+1,p+1))I_{R'T'}\Gamma_R((p+m+1,p+1))= E_{ij}^{(m+p+1)}\, .
$$
Since $\Gamma_R((1,p+m+1))$ will swap the spin in the $(m+p+1)$th site with the spin in site $1$, very similar arguments give
$$
  I_{R'T'}\Gamma_R((1,p+m+1))=\sum_{k=1}^2 E_{ik}^{(1)}E_{kj}^{(m+p+1)},
$$
$$
  \Gamma_R((1,p+m+1))I_{R'T'}=\sum_{k=1}^2 E_{kj}^{(1)}E_{ik}^{(m+p+1)},
$$
$$
  \Gamma_R((1,p+m+1))I_{R'T'}\Gamma_R((1,p+m+1))= E_{ij}^{(m+p+1)}\, .
$$
Our only task now is to evaluate traces of the form
$$
\Tr\Big(\Gamma_R((1,p+m+1))P_{R\to (r,s,t)} I_{R'\, T'} \Gamma_T((1,p+m+1))P_{T\to (u,v,w)}I_{T'\, R'}\Big)\, .
$$
$$
=\sum_{k,l=1}^2 \Tr\Big( E_{ik}^{(1)}E_{kj}^{(m+p+1)}P_{R\to (r,s,t)} E_{jl}^{(1)}E_{li}^{(m+p+1)}P_{T\to (u,v,w)}\Big)
$$
To perform this final trace, our strategy is always the same two steps. For the first step, evaluate the trace over the
$(n+p+1)$th slot. It is clear that the trace over the $p+m+1$th slot factors out and further that
$$
  \Tr (E_{kj}^{(m+p+1)}E_{li}^{(m+p+1)})=\delta_{jl}\delta_{ik}
$$
so that we obtain
$$
  \Tr\Big( E_{ii}^{(1)}P_{R\to (r,s,t)} E_{jj}^{(1)}P_{T\to (u,v,w)}\Big)
$$
To evaluate this final trace we will rewrite the projectors a little. Notice that $E_{kk}^{(1)}$ only has a nontrivial
action on the first site of the spin chain. Thus, we rewrite the projector, separating out the first site. As an example,
consider
$$
  P_{R\to (r,s,t)}=\sum_{\alpha=1}^{d_t} 
\left|j,j^3,\alpha \right\rangle\left\langle j,j^3,\alpha\right|\otimes \sum_{\beta=1}^{d_s}
\left|k,k^3,\beta  \right\rangle\left\langle k,k^3,\beta\right|\, .
$$
To make sense of this formula recall that the labels $j,k,j^3,k^3$ can be traded for the $r,s,t$ labels.
In going from the LHS of this last equation to the RHS we have translated labels and we assure you that
nothing is lost in translation. In figure 2 we remind the reader of how the translation is performed. We
will refer to the Young diagram corresponding to spin $j$, built with $p$ blocks as $s^p_j$ in what follows.
\begin{figure}[h]
          \centering
          {\epsfig{file=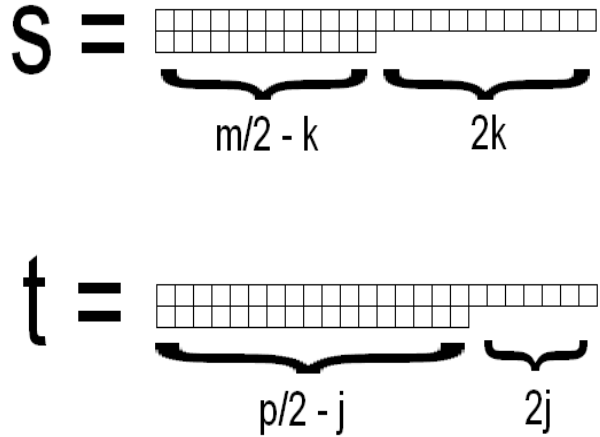,width=8.0cm,height=6.0cm}}
          \caption{How to translate between the $j,k$ and the $s,t$ labels.}
\end{figure}
The piece of the projector that acts on the first $p$ sites is
\begin{equation}
P_{\to t}\equiv \sum_{\alpha=1}^{d_t} \left|j,j^3,\alpha  \right\rangle\left\langle j,j^3,\alpha\right|\, .
\label{reqd}
\end{equation}
If we couple the spins at sites $2,3,...,p$ together, we obtain the states $|j\pm {1\over 2},j^3\pm {1\over 2},\alpha\rangle$ with the
degeneracy label $\alpha$ running from $1$ to the dimension of the irreducible $S_{p-1}$ representation associated to
spin $j\pm {1\over 2}$. This irreducible representation is labeled by the Young diagram $s^{p-1}_{j\pm {1\over 2}}$. The Clebsch-Gordan coefficients
$$
  \left\langle j-{1\over 2},j^{3}-{1\over 2}; {1\over 2},{1\over 2}|j,j^{3}\right\rangle =\sqrt{j+j^3\over 2j}\, ,
$$
$$
  \left\langle j+{1\over 2},j^{3}-{1\over 2}; {1\over 2},{1\over 2}|j,j^{3}\right\rangle =-\sqrt{j-j^3+1\over 2(j+1)}\, ,
$$
$$
  \left\langle j-{1\over 2},j^{3}+{1\over 2}; {1\over 2},-{1\over 2}|j,j^{3}\right\rangle =\sqrt{j-j^3\over 2j}\, ,
$$
$$
  \left\langle j+{1\over 2},j^{3}+{1\over 2}; {1\over 2},-{1\over 2}|j,j^{3}\right\rangle =\sqrt{j+j^3+1\over 2(j+1)}\, .
$$
tell us how to couple the first site with the remaining spins to obtain the projector (\ref{reqd}). Thus, we finally 
have ($s1=s^{p-1}_{j-{1\over 2}}$, $s2=s^{p-1}_{j+{1\over 2}}$)
$$
\left|\phi,\alpha\right\rangle = \sqrt{j+j^3\over 2j} \left|{1\over 2},{1\over 2};j-{1\over 2},j^3-{1\over 2},\alpha  \right\rangle
+\sqrt{j-j^3\over 2j} \left|{1\over 2},-{1\over 2};j-{1\over 2},j^3+{1\over 2},\alpha  \right\rangle\, ,
$$
$$
\left|\psi,\beta\right\rangle = -\sqrt{j-j^3+1\over 2(j+1)} \left|{1\over 2},{1\over 2};j+{1\over 2},j^3-{1\over 2},\beta \right\rangle
+\sqrt{j+j^3+1\over 2(j+1)} \left|{1\over 2},-{1\over 2};j+{1\over 2},j^3+{1\over 2},\beta \right\rangle\, ,
$$
$$
 P_{\to t}= 
\sum_{\alpha=1}^{d_{s1}} \left| \phi,\alpha\right\rangle\left\langle\phi,\alpha\right| +
\sum_{\beta=1}^{d_{s2}}  \left| \psi,\beta\right\rangle\left\langle\psi,\beta\right|\, .
$$
We could of course perform exactly the same manipulations on the projector $P_{\to s}$ that acts on the last $m$ sites of the spin chain.
Now, using the obvious identities
$$
  E_{11}^{(1)}\left|\phi,\alpha\right\rangle = \sqrt{j+j^3\over 2j} \left|{1\over 2},{1\over 2};j-{1\over 2},j^3-{1\over 2},\alpha  \right\rangle\, ,
$$
$$
  E_{22}^{(1)}\left|\phi,\alpha\right\rangle = \sqrt{j-j^3\over 2j} \left|{1\over 2},-{1\over 2};j-{1\over 2},j^3+{1\over 2},\alpha  \right\rangle\, ,
$$
$$
E_{11}^{(1)}\left|\psi,\beta\right\rangle = -\sqrt{j-j^3+1\over 2(j+1)}\left|{1\over 2},{1\over 2};j+{1\over 2},j^3-{1\over 2},\beta \right\rangle\, ,
$$
$$
E_{22}^{(1)}\left|\psi,\beta\right\rangle = \sqrt{j+j^3+1\over 2(j+1)} \left|{1\over 2},-{1\over 2};j+{1\over 2},j^3+{1\over 2},\beta \right\rangle\, ,
$$
it becomes a simple matter to evaluate the above traces. 

Finally, in the limit that we consider, the coefficients of the traces appearing in the dilatation operator are easily evaluated using
$$
{c_{RR'}d_T d_{r'}n\over d_{R'}d_u d_v d_w (n+m+p)}
\sqrt{f_T {\rm hooks}_T{\rm hooks}_r{\rm hooks}_s{\rm hooks}_t\over f_R{\rm hooks}_R{\rm hooks}_u{\rm hooks}_v{\rm hooks}_w}
={\sqrt{c_{RR'}c_{TT'}}\sqrt{{\rm hooks}_s{\rm hooks}_t{\rm hooks}_v{\rm hooks}_w}\over m!p!}\, .
$$
In the above expression, $r'$ is obtained by removing a box from $r$. The box that must be removed from $R$ to obtain $R'$ and the box that
must be removed from $r$ to obtain $r'$ are both removed from the same row. Putting things together we find
\begin{eqnarray}\label{recursion_j3}
 && DO_{j,j^3}(b_0,b_1)=g_{YM}^2 \left[-{1\over 2}\left( m-{(m+2)(j^3)^2\over j(j+1)}\right)
\Delta O_{j,j^3,k,k^3}(b_0,b_1)\right.\nonumber \\
&& + \sqrt{(m + 2j + 4)(m - 2j)\over (2j + 1)(2j+3)} {(j+j^3 + 1)(j-j^3 + 1) \over 2(j + 1)}
 \Delta O_{j+1,j^3,k,k^3}(b_0,b_1)\nonumber \\
&&  +\sqrt{(m + 2j + 2)(m - 2j +2)\over (2j + 1)(2j-1)}
{(j+j^3)(j-j^3)\over 2j} \Delta O_{j-1,j^3,k,k^3}(b_0,b_1)\nonumber\\
&&-{1\over 2}\left( p-{(p+2)(k^3)^2\over k(k+1)}\right)\Delta O_{j,j^3,k,k^3}(b_0,b_1)\nonumber \\
&& + \sqrt{(p + 2k + 4)(p - 2k)\over (2k + 1)(2k+3)} {(k+k^3 +1)(k-k^3 + 1) \over 2(k + 1)}
 \Delta O_{j,j^3,k+1,k^3}(b_0,b_1) \nonumber \\
&& \left. +\sqrt{(p + 2k + 2)(p - 2k +2)\over (2k + 1)(2k-1)}
{(k+k^3 )(k-k^3 ) \over 2 k} \Delta O_{j,j^3,k-1,k^3}(b_0,b_1)
\right]
\end{eqnarray}
where
\begin{eqnarray}
\Delta O(b_0,b_1) &&=\sqrt{(N+b_0)(N+b_0+b_1)}(O(b_0+1,b_1-2)+O(b_0-1,b_1+2))\nonumber \\
 &&-(2N+2b_0+b_1)O(b_0,b_1).\label{}
\end{eqnarray}
Above, we have explicitly carried out the discussion for two long rows. To obtain the result for two long columns, replace
$$
  \sqrt{(N+b_0)(N+b_0+b_1)}\to \sqrt{(N-b_0)(N-b_0-b_1)},\qquad (2N+2b_0+b_1)\to (2N-2b_0-b_1)
$$
in the expression for $\Delta O(b_0,b_1)$. This completes our evaluation of the dilatation operator.

\section{Diagonalization of the Dilatation Operator}

{\sl In this section we reduce the eigenvalue problem for the dilatation operator to the problem of solving a five term recursion
     relation. The explicit solution of this recursion relation allows us to argue that the dilatation operator reduces to a set of
     decoupled oscillators. Thus, the problem we are studying is indeed integrable.}

{\vskip 0.2cm}

We make the following ansatz for the operators of good scaling dimension
$$
  \sum_{b_1}\, f(b_0,b_1)\, O_{pq,j^3,k^3}(b_0,b_1)\,=\sum_{j,k,b_1}\, C_{pq,j^3,k^3}(j,k)\, f(b_0,b_1) O_{j,j^3,k,k^3}(b_0,b_1)\, .
$$
Inserting this ansatz into (\ref{recursion_j3}) we find that the $O_{pq,j^3,k^3}(b_0,b_1)$'s satisfy the recursion relation
\begin{eqnarray}
&& -\alpha_{rq,j^3,k^3}C_{rq,j^3,k^3}(j,k)=
\sqrt{(m + 2j + 4)(m - 2j)\over (2j + 1)(2j+3)} {(j+j^3 + 1)(j-j^3 + 1) \over 2(j + 1)} C_{rq,j^3,k^3}(j+1,k)\nonumber \\
&&+\sqrt{(m + 2j + 2)(m - 2j +2)\over (2j + 1)(2j-1)} {(j+j^3 )(j-j^3 ) \over 2 j}C_{rq,j^3,k^3}(j-1,k)\nonumber\\
&&-{1\over 2}\left( m-{(m+2)(j^3)^2\over j(j+1)}\right)C_{rq,j^3,k^3}(j,k)\, \nonumber \\
&&+\sqrt{(p + 2k + 4)(p - 2k)\over (2k + 1)(2k+3)} {(k+k^3 + 1)(k-k^3 + 1) \over 2(k + 1)} C_{rq,j^3,k^3}(j,k+1) \nonumber \\
&&+\sqrt{(p + 2k + 2)(p - 2k +2)\over (2k + 1)(2k-1)} {(k+k^3 )(k-k^3 ) \over 2 k}C_{rq,j^3,k^3}(j,k-1)\nonumber\\
&&-{1\over 2}\left( p-{(p+2)(k^3)^2\over k(k+1)}\right)C_{rq,j^3,k^3}(j,k)\, .\nonumber \\
\end{eqnarray}
Exploiting the $j^3 \to -j^3$ and $k^3\to -k^3$ symmetries of this equation, we need only solve for the $j^3 \ge 0$ and $k^3 \ge 0$
cases. The ranges for $j$ and $k$ are
\begin{eqnarray}
0 \le |j^3|\le j\le {m\over 2}\qquad 0 \le |k^3|\le k\le {p\over 2}\, . \nonumber
\end{eqnarray}
From the form of the recursion relation, it is natural to make the ``separation of variables'' ansatz
$$
  C_{rq,j^3,k^3}(j,k)=C_{r,j^3}(j)C_{q,k^3}(k)\, .
$$
Our five term recurrence relation now reduces to two three term recurrence relations 
\begin{eqnarray}
&& -\alpha_{r,j^3}C_{p,j^3}(j,)=
\sqrt{(m + 2j + 4)(m - 2j)\over (2j + 1)(2j+3)} {(j+j^3 + 1)(j-j^3 + 1) \over 2(j + 1)} C_{r,j^3}(j+1) \\
&&+\sqrt{(m + 2j + 2)(m - 2j +2)\over (2j + 1)(2j-1)} {(j+j^3 )(j-j^3 ) \over 2 j}C_{r,j^3}(j-1)
-{1\over 2}\left( m-{(m+2)(j^3)^2\over j(j+1)}\right)C_{r,j^3}(j)\, ,\nonumber
\end{eqnarray}
\begin{eqnarray}
&& -\alpha_{q,k^3}C_{q,k^3}(k)=
\sqrt{(p + 2k + 4)(p - 2k)\over (2k + 1)(2k+3)} {(k+k^3 + 1)(k-k^3 + 1) \over 2(k + 1)} C_{q,k^3}(k+1) \\
&&+\sqrt{(p + 2k + 2)(p - 2k +2)\over (2k + 1)(2k-1)} {(k+k^3 )(k-k^3 ) \over 2 k}C_{q,k^3}(k-1)
-{1\over 2}\left( p-{(p+2)(k^3)^2\over k(k+1)}\right)C_{q,k^3}(k)\, .\nonumber 
\end{eqnarray}
These are identical to the three term recursion relations that appear in \cite{Carlson:2011hy}.
To solve these recurrence relations, introduce the Hahn polynomial\cite{hahnhelp}
$$
Q_n(x;\alpha,\beta,N)\equiv 
{}_3F_2\left({}^{-n,n+\alpha+\beta+1,-x}_{\alpha+1,-N}\Big|1\right)
$$
From the recurrence relation obeyed by Hahn polynomials (see equation (1.5.3) in \cite{hahnhelp}) we have
{\small
$$
r\,{}_3F_2\left({}^{|j^3|-j,j+1+|j^3|,-r}_{1,|j^3|-{m\over 2}}\Big|1\right)=
{(j+j^3 + 1)(j-j^3 + 1)(m-2j) \over 2(j + 1)(2j+1)}
{}_3F_2\left({}^{-1+|j^3|-j,j+2+|j^3|,-r}_{1,|j^3|-{m\over 2}}\Big|1\right)
$$
$$
-\left( {m\over 2}-{(m+2)(j^3)^2\over 2j(j+1)}\right)
{}_3F_2\left({}^{|j^3|-j,j+1+|j^3|,-r}_{1,|j^3|-{m\over 2}}\Big|1\right)
+{(j+j^3 )(j-j^3 )(m+2j+2) \over 2 j(2j+1)}
{}_3F_2\left({}^{1+|j^3|-j,j+|j^3|,-r}_{1,|j^3|-{m\over 2}}\Big|1\right)
$$
}
Consequently, our recursion relation is solved by
\begin{equation}
C_{r,j^3}(j)=(-1)^{{m\over 2}-p}\left({m\over 2}\right)!\sqrt{(2j+1)\over \left({m\over 2}-j\right)!\left({m\over 2}+j+1\right)!}
{}_3F_2\left({}^{|j^3|-j,j+|j^3|+1,-r}_{|j^3|-{m\over 2},1}\Big| 1\right)
\end{equation}
$$
|j^3| \le j\le {m\over 2},\qquad 0\le r\le {m\over 2}-|j^3|
$$
and
\begin{equation}
C_{q,k^3}(k)=(-1)^{{p\over 2}-q}\left({p\over 2}\right)!\sqrt{(2k+1)\over \left({p\over 2}-k\right)!\left({p\over 2}+k+1\right)!}
{}_3F_2\left({}^{|k^3|-k,k+|k^3|+1,-q}_{|k^3|-{p\over 2},1}\Big| 1\right)
\end{equation}
$$
|k^3| \le k\le {p\over 2},\qquad 0\le q\le {p\over 2}-|k^3|\, .
$$
The associated eigenvalues are
$$ 
-\alpha_{rq,j^3,k^3} = -2(r+q) = 0,-2,-4,...,-(m-2|j^3|+p-2|k^3|)\, .
$$
Our eigenfunctions are essentially the Hahn polynomials. It is a well known fact that the Hahn polynomials are closely
related to the Clebsch-Gordan coefficients of $SU(2)$ \cite{HP}.

The eigenproblem of the dilatation operator now reduces to solving
$$
  \lambda \sum_{b_1}\, f(b_0,b_1)\, O_{rq,j^3,k^3}(b_0,b_1)
 =-\alpha_{rq,j^3,k^3}\sum_{b_1}\, f(b_0,b_1)\, \Delta O_{rq,j^3,k^3}(b_0,b_1)\, .
$$
This eigenproblem implies $f(b_0,b_1)$ satisfy the recursion relation
\begin{eqnarray}
&&-\alpha_{rq,j^3,k^3} g_{YM}^{2}[\sqrt{(N+b_0)(N+b_0+b_1)}(f(b_0-1,b_1+2)+f(b_0+1,b_1-2))\nonumber \\
&&-(2N+2b_0+b_1)f(b_0,b_1)]=\lambda f(b_0, b_1)
\label{Kp}
\end{eqnarray}
Since we work at large $N$, we can replace (\ref{Kp}) by
\begin{eqnarray}
&&\lambda f(b_0, b_1) = -\alpha_{rq,j^3,k^3} g_{YM}^{2}[\sqrt{(N+b_0)(N+b_0+b_1+1)}f(b_0-1,b_1+2)\nonumber \\
&&+\sqrt{(N+b_0+1)(N+b_0+b_1)}f(b_0+1,b_1-2)
-(2N+2b_0+b_1)f(b_1,b_1)]\, .
\nonumber
\end{eqnarray}
This recursion relation is precisely the recursion relation of the finite oscillator \cite{finiteoscillator}!
In the continuum limit (which corresponds to the large $N$ limit) we recover the usual description of the harmonic
oscillator, demonstrating rather explicitly that the eigenproblem of the dilatation operator reduces to solving a set
of decoupled harmonic oscillators. The solution to (\ref{Kp}) is \cite{finiteoscillator}
\begin{equation}
f(b_0,b_1)={(-1)^n  ({1\over2})^{N+b_0+{b_1 \over
2}}}\sqrt{\left(^{2N+2b_0+b_1}_{N+b_0+b_1}\right)\left(^{2N+2b_0+b_1}_{\quad\quad
n}\right)}{}_2F_1({}^{-n,-(N+b_0+b_1)}_{-(2N+2b_0+b_1)}\Big| 2)\, .
\label{alllevels}
\end{equation}
These solutions are closely related to the symmetric Kravchuk polynomial $K_n(x,1/q,p)$ defined by
$$
 {}_2F_1\left({}^{-n,-x}_{-p};q\right)=K_n(x,1/q,p)\, .
$$
The corresponding eigenvalue is $\lambda = 2n\alpha_{rq,j^3,k^3} g_{YM}^2$. Recall that $b_1 \ge 0$ so that only half of 
the wavefunctions are selected (those that vanish when $b_1=0$) and consequently the eigenvalue $\lambda$ level spacing is
$4\alpha_{rq,j^3,k^3} g_{YM}^2 = 8(p+q)g_{YM}^2$. 

\section{Discussion}

In this article we have studied the action of the dilatation operator on restricted Schur polynomials $\chi_{R,(r,s,t)}(Z,Y,X)$,
built from three complex scalars $X$, $Y$ and $Z$ and labeled by Young diagrams with at most two rows or two columns. The 
operators have $O(N)$ fields of each of the three flavors, but there are many many more $Z$s than $X$s or $Y$s. Our main
result is that the dilatation operator reduces to a set of decoupled oscillators and is hence an integrable system. If
we have $m$ $Y$s and $p$ $X$s with $p,m$ both even, we obtain a set of oscillators with frequency $\omega_{ij}$ and
degeneracy $d_{ij}$ given by
$$
  \omega_{ij}=8(i+j)g_{YM}^2,\qquad d_{ij}=\left( 2(m-i)+1\right)\left( 2(p-j) +1\right),
$$
$$
     i=0,1,...,m,\qquad j=0,1,...,p\, .
$$
If $p$ is even and $m$ is odd we have
$$
  \omega_{ij}=8(i+j)g_{YM}^2,\qquad d_{ij}=2 \left(m-i+1\right)\left( 2(p-j) +1\right),
$$
$$
     i=0,1,...,m,\qquad j=0,1,...,p\, .
$$
If $m$ is even and $p$ is odd we have
$$
  \omega_{ij}=8(i+j)g_{YM}^2,\qquad d_{ij}=2 \left(2(m-i)+1\right)\left( p-j +1\right),
$$
$$
     i=0,1,...,m,\qquad j=0,1,...,p\, .
$$
If both $p$ and $m$ are odd we have
$$
  \omega_{ij}=8(i+j)g_{YM}^2,\qquad d_{ij}=4 \left(m-i+1\right)\left( p-j +1\right),
$$
$$
     i=0,1,...,m,\qquad j=0,1,...,p\, .
$$
The oscillators corresponding to a zero frequency are BPS operators built using three complex scalars $X$, $Y$ and $Z$.

The form of the dilatation operator (\ref{recursion_j3}) is intriguing: it looks like the sum of two of the dilatation
operators computed in \cite{Carlson:2011hy}, with one acting on the $Y$s (with quantum numbers $k,k^3$) and one
acting on the $X$s (with quantum numbers $j,j^3$). With the benefit of hindsight, could we have anticipated this
structure? The bulk of our effort involved evaluating traces like this one
$$
 \Tr\Big(\Big[ \Gamma_R((p+m+1,p+1)),P_{R\to (r,s,t)}\Big]I_{R'\, T'}\Big[\Gamma_T((p+m+1,p+1)),P_{T\to (u,v,w)}\Big]I_{T'\, R'}\Big)\, .
$$
Notice that both $\Gamma_R((p+m+1,p+1))$ and $I_{R'\, T'}$ do not act on the first $p$ sites of the spin chain. Further,
our projector factorizes into a projector acting on the first $p$ sites times a projector acting on the remaining $m$ sites.
Consequently, the trace over the first $p$ sites gives $\delta_{tw}d_w$. The trace that remains is exactly of the form
considered in \cite{Carlson:2011hy}, explaining our final answer (\ref{recursion_j3}). An important new feature we have found 
here, described in detail in Appendix C, is that before making the approximations described in section 3.1, the spectrum of
the dilatation operator is not equivalent to a collection of harmonic oscillators. This is similar to what one finds in
the sector of operators with a bare dimension of order $O(1)$: in the large $N$ limit (which in this case is the planar limit)
one obtains an integrable system. Adding $1/N$ corrections seems to spoil the integrability \cite{Kristjansen:2010kg,Zoubos:2010kh}.

Apart from computing the spectrum of the dilatation operator, we have managed to compute the associated eigenstates.
These states are given in terms of Kravchuk polynomials and Hahn polynomials. The Hahn polynomials are closely related
to the wave functions of the one dimensional harmonic oscillator\cite{finiteoscillator} while the Hahn polynomials are
closely related to the wave functions of the 2d radial oscillator\cite{Carlson:2011hy}. The argument of these polynomials
are given by $j$, $k$ or $b_1$, which have a direct link to the Young diagrams labeling the operators, as summarized for
example in figure 2\footnote{The Young diagram $r$ is not shown in figure 2. The number of columns with a single box is
given by $b_1$.}. Thus, the ``space'' on which the wave functions are defined comes from the Young diagram itself.
Based on our experience with the half BPS sector, it is natural to associate each one of the rows of the Young diagram 
with each one of the giant gravitons. Recalling that $Y=\phi_3+i\phi_4$ we know that the number of $Y$s in each operator 
tells us the angular momentum of the operator in the 3-4 plane. Similarly, the number of $X$s in each operator 
tells us the angular momentum of the operator in the 5-6 plane and the number of $X$s in each operator tells us the angular 
momentum of the operator in the 1-2 plane. Giving an angular momentum to the giant gravitons will cause them to expand as a 
consequence of the Myers effect\cite{Myers:1999ps}. Thus, for example, the separation between the two gravitons in the 3-4 
plane will be related to the difference in angular momenta of the two giants. Consequently, the quantum number $k$ is acting
like a coordinate for the radial separation between the two giants in the 3-4 plane. Thus, we see very concretely the
emergence of local physics from the system of Young diagrams labeling the restricted Schur polynomial. This is
strongly reminiscent of the 1/2 BPS case where the Schur polynomials provide wave functions for fermions in a harmonic
oscillator potential and further, these wave functions very naturally reproduce features of the geometries 
and the phase space \cite{Lin:2004nb}.

For the matrix model we are studying here it is not true that the matrices $Z$,$Y$,$X$ commute, we can't simultaneously diagonalize them 
and there is no analog of the eigenvalue basis that is so useful for the large $N$ dynamics of single matrix models. For the subsystem
describing the BPS states however \cite{Berenstein:2005aa} has argued that the matrices might commute in the interacting 
theory and hence there may be a description in terms of eigenvalues. The argument uses the fact that the weak coupling and
strong coupling limits of the BPS sector agree and the fact that at strong coupling we can be confident that the matrices commute.
If this is the case, the eigenvalue dynamics should be the dynamics in an oscillator potential with repulsions preventing the collision 
of eigenvalues. We have described a part of the BPS sector (as well as non-BPS operators) among the operators we have studied. We do indeed
find the dynamics of harmonic oscillators. In the case of a single matrix it is possible to associate the rows of the Young
diagram labeling a Schur polynomial with the eigenvalues of the matrix\cite{Koch:2008ah}. This provides a connection between the eigenvalue
description and the Schur polynomial description for single matrix models. Our results suggest this might have a generalization 
to multimatrix models.

The operators we have considered are dual to giant gravitons. A connection between the geometry of giant gravitons and harmonic oscillators 
was already uncovered in \cite{Biswas:2006tj,Mandal:2006tk,jurgis}. This work quantizes the moduli space of Mikhailov's giant gravitons so
that one is capturing a huge space of states. It is this huge space of states that connects to harmonic oscillators. Our study is focused 
on a two giant system. Consequently, the oscillators that we have found are associated to this two giant system and excitations of it. It is
natural to think that our oscillators arise from the quantization of the possible excitation modes of a giant graviton.

{\vskip 0.2cm}

{\vskip 1.0cm}

\noindent
{\it Acknowledgements:}
We would like to thank Dimitrios Giataganas, Norman Ives, Sanjaye Ramgoolam and Michael Stephanou  
for pleasant discussions and/or helpful correspondence. This work is based upon research supported by the 
South African Research Chairs Initiative of the Department of Science and Technology and National Research 
Foundation. Any opinion, findings and conclusions or recommendations expressed in this material are those of 
the authors and therefore the NRF and DST do not accept any liability with regard thereto.

\appendix

\section{Example Projector}
\label{projector}

In the section we will consider the case that $m=p=3$. Towards this end, we couple the states of 3 spin 
${1\over 2}$-particles to obtain
$$
  \left| -{1\over 2},-{1\over 2},-{1\over 2}\right\rangle =\left| {3\over 2},-{3\over 2}\right\rangle\, ,
$$
$$
  \left| -{1\over 2},-{1\over 2}, {1\over 2}\right\rangle ={1\over \sqrt{2}}\left| {1\over 2},-{1\over 2}\right\rangle^A
  +{1\over\sqrt{6}}\left| {1\over 2},-{1\over 2}\right\rangle^B +{1\over\sqrt{3}}\left| {3\over 2},-{1\over 2}\right\rangle\, ,
$$
$$
  \left| -{1\over 2}, {1\over 2},-{1\over 2}\right\rangle =-{1\over \sqrt{2}}\left| {1\over 2},-{1\over 2}\right\rangle^A
  +{1\over\sqrt{6}}\left| {1\over 2},-{1\over 2}\right\rangle^B +{1\over\sqrt{3}}\left| {3\over 2},-{1\over 2}\right\rangle\, ,
$$
$$
\left| -{1\over 2}, {1\over 2}, {1\over 2}\right\rangle =
  +\sqrt{2\over 3}\left| {1\over 2}, {1\over 2}\right\rangle^B +{1\over\sqrt{3}}\left| {3\over 2}, {1\over 2}\right\rangle\, ,
$$
$$
  \left| {1\over 2},-{1\over 2},-{1\over 2}\right\rangle =
  -\sqrt{2\over 3}\left| {1\over 2},-{1\over 2}\right\rangle^B +{1\over\sqrt{3}}\left| {3\over 2},-{1\over 2}\right\rangle\, ,
$$
$$
\left|  {1\over 2},-{1\over 2}, {1\over 2}\right\rangle ={1\over \sqrt{2}}\left| {1\over 2},{1\over 2}\right\rangle^A
  -{1\over\sqrt{6}}\left| {1\over 2},{1\over 2}\right\rangle^B +{1\over\sqrt{3}}\left| {3\over 2},{1\over 2}\right\rangle\, ,
$$
$$
  \left| {1\over 2},{1\over 2},-{1\over 2}\right\rangle =-{1\over \sqrt{2}}\left| {1\over 2}, {1\over 2}\right\rangle^A
  -{1\over\sqrt{6}}\left| {1\over 2}, {1\over 2}\right\rangle^B +{1\over\sqrt{3}}\left| {3\over 2}, {1\over 2}\right\rangle\, ,
$$
$$
\left| {1\over 2},{1\over 2},{1\over 2}\right\rangle =\left| {3\over 2},{3\over 2}\right\rangle
$$
The spin ${3\over 2}$ representation is organized by $S_3$ irreducible representation $\yng(3)$, which is one dimensional, so that
the spin ${3\over 2}$ multiplet is not degenerate. The spin ${1\over 2}$ representation is organized by $S_3$ irreducible representation
$\yng(2,1)$ which is two dimensional. Consequently, the spin ${1\over 2}$ occurs with degeneracy 2. $A$ and $B$ label the two multiplets.
Thus, picking a particular state, $A$ and $B$ should label the two states in the $S_3$ irreducible representation which is labeled
by the Young diagram $\yng(2,1)$. From the results above we easily find
$$
  \left| {1\over 2}, {1\over 2}\right\rangle^A = {1\over\sqrt{2}}\left|  {1\over 2},-{1\over 2}, {1\over 2}\right\rangle
                                                -{1\over\sqrt{2}}\left|  {1\over 2}, {1\over 2},-{1\over 2}\right\rangle
$$
$$
  \left| {1\over 2}, {1\over 2}\right\rangle^B = -{1\over\sqrt{6}}\left|  {1\over 2},-{1\over 2}, {1\over 2}\right\rangle
                                                 -{1\over\sqrt{6}}\left|  {1\over 2}, {1\over 2},-{1\over 2}\right\rangle
                                                 +\sqrt{2\over 3} \left| -{1\over 2}, {1\over 2}, {1\over 2}\right\rangle
$$
Taking the direct product with another such multiplet arising from coupling a further three spins, we should obtain the four states of the 
$S_3\times S_3$ irreducible representation labeled by the pair of Young diagrams $(\yng(2,1),\yng(2,1))$. These four states are easily 
constructed
$$
  \left|1,1\right\rangle ={1\over 2}\left|{1\over 2},-{1\over 2}, {1\over 2},{1\over 2},-{1\over 2}, {1\over 2}\right\rangle
                         -{1\over 2}\left|{1\over 2},-{1\over 2}, {1\over 2},{1\over 2}, {1\over 2},-{1\over 2}\right\rangle
                         -{1\over 2}\left|{1\over 2}, {1\over 2},-{1\over 2},{1\over 2},-{1\over 2}, {1\over 2}\right\rangle
$$
$$
                         +{1\over 2}\left|{1\over 2}, {1\over 2},-{1\over 2},{1\over 2}, {1\over 2},-{1\over 2}\right\rangle
$$

$$
  \left|1,2\right\rangle =-{1\over 2\sqrt{3}}\left|{1\over 2},-{1\over 2}, {1\over 2},{1\over 2},-{1\over 2}, {1\over 2}\right\rangle
                          -{1\over 2\sqrt{3}}\left|{1\over 2}, {1\over 2},-{1\over 2},{1\over 2},-{1\over 2}, {1\over 2}\right\rangle
                          +{1\over  \sqrt{3}}  \left|{1\over 2},-{1\over 2}, {1\over 2},-{1\over 2},{1\over 2}, {1\over 2}\right\rangle
$$
$$
                         +{1\over 2\sqrt{3}}\left|{1\over 2}, {1\over 2},-{1\over 2},{1\over 2},-{1\over 2}, {1\over 2}\right\rangle
                         +{1\over 2\sqrt{3}}\left|{1\over 2}, {1\over 2},-{1\over 2},{1\over 2}, {1\over 2},-{1\over 2}\right\rangle
                         -{1\over  \sqrt{3}}\left|{1\over 2}, {1\over 2},-{1\over 2},-{1\over 2},{1\over 2}, {1\over 2}\right\rangle
$$

$$
  \left|2,1\right\rangle =-{1\over 2\sqrt{3}}\left|{1\over 2},-{1\over 2}, {1\over 2},{1\over 2},-{1\over 2}, {1\over 2}\right\rangle
                          -{1\over 2\sqrt{3}}\left|{1\over 2}, {1\over 2},-{1\over 2},{1\over 2},-{1\over 2}, {1\over 2}\right\rangle
                          +{1\over  \sqrt{3}}\left|-{1\over 2},{1\over 2}, {1\over 2},{1\over 2},-{1\over 2}, {1\over 2}\right\rangle
$$
$$
                         +{1\over 2\sqrt{3}}\left|{1\over 2},-{1\over 2}, {1\over 2},{1\over 2}, {1\over 2},-{1\over 2}\right\rangle
                         +{1\over 2\sqrt{3}}\left|{1\over 2}, {1\over 2},-{1\over 2},{1\over 2}, {1\over 2},-{1\over 2}\right\rangle
                         -{1\over  \sqrt{3}}\left|-{1\over 2},{1\over 2}, {1\over 2},{1\over 2}, {1\over 2},-{1\over 2}\right\rangle
$$

$$
  \left|2,2\right\rangle ={1\over 6}\left|{1\over 2},-{1\over 2}, {1\over 2},{1\over 2},-{1\over 2}, {1\over 2}\right\rangle
                         +{1\over 6}\left|{1\over 2},-{1\over 2}, {1\over 2},{1\over 2}, {1\over 2},-{1\over 2}\right\rangle
                         -{1\over 3}\left|{1\over 2},-{1\over 2}, {1\over 2},-{1\over 2},{1\over 2}, {1\over 2}\right\rangle
$$
$$
                         +{1\over 6}\left|{1\over 2}, {1\over 2},-{1\over 2},{1\over 2},-{1\over 2}, {1\over 2}\right\rangle
                         +{1\over 6}\left|{1\over 2}, {1\over 2},-{1\over 2},{1\over 2}, {1\over 2},-{1\over 2}\right\rangle
                         -{1\over 3}\left|{1\over 2}, {1\over 2},-{1\over 2},-{1\over 2},{1\over 2}, {1\over 2}\right\rangle
$$
$$
                         -{1\over 3}\left|-{1\over 2},{1\over 2}, {1\over 2},{1\over 2},-{1\over 2}, {1\over 2}\right\rangle
                         -{1\over 3}\left|-{1\over 2},{1\over 2}, {1\over 2},{1\over 2}, {1\over 2},-{1\over 2}\right\rangle
                         +{2\over 3}\left|-{1\over 2},{1\over 2}, {1\over 2},-{1\over 2},{1\over 2}, {1\over 2}\right\rangle
$$

It is rather simple to check that these four states do indeed span the carrier space of the $S_3\times S_3$ representation
labeled by $(\yng(2,1),\yng(2,1))$. As an example, $(12)$ has a matrix representation
$$
\Gamma(12)=\left[
\matrix{{1\over 2}        &{\sqrt{3}\over 2} &0                 &0\cr
        {\sqrt{3}\over 2} &-{1\over 2}       &0                 &0\cr
        0                 &0                 &{1\over 2}        &{\sqrt{3}\over 2}\cr
        0                 &0                 &{\sqrt{3}\over 2} &-{1\over 2}}
\right]=
\left[
\matrix{{1\over 2}        &{\sqrt{3}\over 2}\cr
        {\sqrt{3}\over 2} &-{1\over 2}      }
\right]\otimes
\left[
\matrix{1        &0  \cr
        0        &1      }
\right]=\Gamma_{\yng(2,1)}\left( (12)\right)\otimes \Gamma_{\yng(2,1)}\left( {\bf 1}\right)\, .
$$
Given a basis of the required carrier space, it is now trivial to construct the associated projector.

\section{The Space $L(\Omega_{m,p})$}
\label{rep}
In this Appendix we discuss the representation theory relevant for this article.
We highly recommend the article \cite{GelPair} for related background material.
Consider the group $S_p\times S_m$. Define
$$
\Omega_{k,l}= (S_p/S_{p-l}\times S_l)\times (S_m/S_{m-k}\times S_k)
$$
to be the space of all pairs of $k,l$ subsets, where the $k$ subsets are subsets of $\{1,2,...,p\}$ and the
$l$ subsets are subsets of $\{p+1,p+2,...,p+m\}$. If $p=2$ and $m=2$ then $\Omega_{1,1}=\{\{1;3\},\{1;4\},\{2;3\},\{2;4\}\}$ and
$\Omega_{2,2}=\{\{1,2;3,4\}\}$ etc. You can identify a $k,l$ subset with a monomial.
For example, we'd identify $\{1;3\}$ with $x_1 y_3$ and $\{1,2;4\}$ with $x_1 x_2 y_4$. Thus, we can consider
$\Omega_{k,l}$ to be the space of distinct monomials in two types of variables ($x_i$ and $y_i$) with $k+l$ factors and no factor repeats. 
Ordering of the factors is not important so that $x_1 x_2 y_4$ and $y_4 x_1 x_2$ are exactly the same element of $\Omega_{2,1}$. Our main 
interest is in $L(\Omega_{k,l})$ which is the space of complex valued functions on $\Omega_{k,l}$. The group $S_p\times S_m$ has a very 
natural action on $L(\Omega_{k,l})$: we can define this action by defining it on each monomial.
The symmetric group $S_m\subset S_p\times S_m$ acts by permuting the labels on the $x_i$ factors in the monomial and
the symmetric group $S_p\subset S_p\times S_m$ acts by permuting the labels on the $y_i$ factors in the monomial. Thus, for example,
for $m=3=p$
$$ (12)x_1 x_2 y_4 = x_1 x_2 y_4 \qquad
   (45)x_1 x_2 y_4 = x_1 x_2 y_5 \, .
$$
There is a natural inner product under which the monomials are orthonormal, so that, for example
$$
\langle x_1 x_2 y_4,x_1 x_2 y_4\rangle =1,\qquad \langle x_1 x_2 y_4 ,x_1 x_3 y_4\rangle =0= \langle x_1 x_2 y_4,x_1 x_2 y_5\rangle\, .
$$
$L(\Omega_{k,l})$ furnishes a reducible representation of the group $S_m\times S_p$. The relevance of $L(\Omega_{k,l})$ for us here
is that the projectors acting in $L(\Omega_{k,l})$ projecting onto an irreducible representation of $S_p\times S_m$ are precisely the
projectors we need to define the restricted Schur polynomials. Consider the operator
\begin{equation}\label{}
d_1=\sum_{i=1}^p {\partial\over \partial x_i}\, .
\end{equation}
It maps from $L(\Omega_{k,l})$ to $L(\Omega_{k-1,l})$. Further, it commutes with the action of $S_p\times S_m$. Because
of this, elements of the kernel of $d_1$ form an invariant $S_p\times S_m$ subspace. Similarly,
\begin{equation}\label{}
d_2=\sum_{i=p+1}^{p+m} {\partial\over \partial y_i}\, ,
\end{equation}
maps $L(\Omega_{k,l})$ to $L(\Omega_{k,l-1})$ and it also commutes with the action of $S_p\times S_m$. Thus, the elements of the
kernel of $d_2$ will also form an invariant $S_p\times S_m$ subspace. Using results from \cite{GelPair} it follows that the intersection
of the kernel of $d_1$, the kernel of $d_2$ and $L(\Omega_{k,l})$ is an irreducible representation of $S_p\times S_m$.

An example will help to make this discussion concrete. For $m=3=p$ the intersection of the kernel of $d_1$, the kernel of $d_2$ and
$L(\Omega_{1,1})$ is clearly spanned by the polynomials
$$
\phi_1={x_1-x_2\over\sqrt{2}}{y_4-y_5\over\sqrt{2}},\qquad
\phi_2={x_1-x_2\over\sqrt{2}}{y_4+y_5-2y_6\over\sqrt{6}},
$$
$$
\phi_3={x_1+x_2-2x_3\over\sqrt{6}}{y_4-y_5\over\sqrt{2}},\qquad
\phi_4={x_1+x_2-2x_3\over\sqrt{6}}{y_4+y_5-2y_6\over\sqrt{6}}\, .
$$
It is easy to check that
$$
(12)\phi_1=-\phi_1,\qquad
(12)\phi_2=-\phi_2,\qquad
(12)\phi_3= \phi_3,\qquad
(12)\phi_4= \phi_4,
$$
$$
(23)\phi_1= {1\over 2}\phi_1+{\sqrt{3}\over 2}\phi_3,\qquad
(23)\phi_2= {1\over 2}\phi_2+{\sqrt{3}\over 2}\phi_4,
$$
$$
(23)\phi_3= -{1\over 2}\phi_3+{\sqrt{3}\over 2}\phi_1,\qquad
(23)\phi_4= -{1\over 2}\phi_4+{\sqrt{3}\over 2}\phi_2,
$$
$$
(45)\phi_1=-\phi_1,\qquad
(45)\phi_2= \phi_2,\qquad
(45)\phi_3=-\phi_3,\qquad
(45)\phi_4= \phi_4,
$$
$$
(56)\phi_1= {1\over 2}\phi_1+{\sqrt{3}\over 2}\phi_2,\qquad
(56)\phi_2=-{1\over 2}\phi_2+{\sqrt{3}\over 2}\phi_1,
$$
$$
(56)\phi_3=  {1\over 2}\phi_3+{\sqrt{3}\over 2}\phi_4,\qquad
(56)\phi_4= -{1\over 2}\phi_4+{\sqrt{3}\over 2}\phi_3,
$$
Thus, we have the following group elements
$$
\Gamma\left( (12)\right)=
\left[\begin {array}{cccc} -1&0&0&0\\
\noalign{\medskip}0&-1&0&0\\
\noalign{\medskip}0&0&1&0\\
\noalign{\medskip}0&0&0&1\end {array} \right]
=
\left[\begin {array}{cc} -1&0\\
\noalign{\medskip}0&1\end {array} \right]\otimes
\left[\begin {array}{cc} 1&0\\
\noalign{\medskip}0&1\end {array} \right],
$$
$$
\Gamma\left( (23)\right)=
\left[\begin {array}{cccc} {1\over 2} &0                 &{\sqrt{3}\over 2}      &0                \\
\noalign{\medskip}              0     &{1\over 2}        &0                      &{\sqrt{3}\over 2}\\
\noalign{\medskip}   {\sqrt{3}\over 2}&0                 &-{1\over 2}            &0                \\
\noalign{\medskip}              0     &{\sqrt{3}\over 2} &0                      &-{1\over 2}        \end {array} \right]
=
\left[\begin {array}{cc} {1\over 2}        & {\sqrt{3}\over 2}\\
\noalign{\medskip}       {\sqrt{3}\over 2} &-{1\over 2}    \end {array} \right]\otimes
\left[\begin {array}{cc} 1&0\\
\noalign{\medskip}0&1\end {array} \right],
$$
$$
\Gamma\left( (45)\right)=
\left[\begin {array}{cccc} -1&0&0&0\\
\noalign{\medskip}0&1&0&0\\
\noalign{\medskip}0&0&-1&0\\
\noalign{\medskip}0&0&0&1\end {array} \right]
=
\left[\begin {array}{cc} 1&0\\
\noalign{\medskip}0&1\end {array} \right]\otimes
\left[\begin {array}{cc} -1&0\\
\noalign{\medskip}0&1\end {array} \right],
$$
$$
\Gamma\left( (56)\right)=
\left[\begin {array}{cccc} {1\over 2}            &{\sqrt{3}\over 2} &0                      &0                \\
\noalign{\medskip}         {\sqrt{3}\over 2}     &-{1\over 2}       &0                      &0                \\
\noalign{\medskip}   {\sqrt{3}\over 2}           &0                 &{1\over 2}             &{\sqrt{3}\over 2}\\
\noalign{\medskip}              0                &0                 &{\sqrt{3}\over 2}      &-{1\over 2}        \end {array} \right]
=
\left[\begin {array}{cc} 1&0\\
\noalign{\medskip}0&1\end {array} \right]
\otimes
\left[\begin {array}{cc} {1\over 2}        & {\sqrt{3}\over 2}\\
\noalign{\medskip}       {\sqrt{3}\over 2} &-{1\over 2}    \end {array} \right]\, .
$$
Using these matrices it is possible to compute all elements of the group now, and then to compute characters. In this way,
it is a simple matter to identify this as the $(\yng(2,1),\yng(2,1))$ irreducible representation of $S_3\times S_3$.

\section{Explicit Evaluation of the Dilatation Operator for $m=p=2$ and Numerical Spectrum}

We have explicitly evaluated the dilatation operator (\ref{evlaute}) for the case $m=p=2$. There are a total of 16 operators
that can be defined. Our notation for these operators is $O_{R,(r,s,t)}=O_i(b_0,b_1)$. The labels $b_0$ and $b_1$ specifies the 
second label of the restricted Schur polynomial: $r$ has $b_0$ rows with two boxes and $b_1$ rows with a single box. The label
$i=1,...,16$ tells you what the labels $s,t$ are and it tells you how the boxes are removed from $R$ to obtain $r$. These labels
are defined as 
$$
   O_1=O_{\tiny\yng(2,2,2,2,2,2,2,1,1,1,1,1,1,1),\,\yng(2,2,2,2,2,2,2,1,1,1)\,\yng(1,1)\,\yng(1,1)}\quad
   O_2=O_{\tiny\yng(2,2,2,2,2,2,2,1,1,1,1,1,1,1),\,\yng(2,2,2,2,2,2,1,1,1,1,1)\,\yng(1,1)\,\yng(1,1)}\quad
   O_3=O_{\tiny\yng(2,2,2,2,2,2,2,1,1,1,1,1,1,1),\,\yng(2,2,2,2,2,2,1,1,1,1,1)\,\yng(1,1)\,\yng(2)}\quad
   O_4=O_{\tiny\yng(2,2,2,2,2,2,2,1,1,1,1,1,1,1),\,\yng(2,2,2,2,2,1,1,1,1,1,1,1)\,\yng(1,1)\,\yng(1,1)}
$$
$$
   O_5=O_{\tiny\yng(2,2,2,2,2,2,2,1,1,1,1,1,1,1),\,\yng(2,2,2,2,2,2,1,1,1,1,1)\,\yng(1,1)\,\yng(1,1)}\quad
   O_6=O_{\tiny\yng(2,2,2,2,2,2,2,1,1,1,1,1,1,1),\,\yng(2,2,2,2,2,2,1,1,1,1,1)\,\yng(2)\,\yng(1,1)}\quad
   O_7=O_{\tiny\yng(2,2,2,2,2,2,2,1,1,1,1,1,1,1),\,\yng(2,2,2,2,1,1,1,1,1,1,1,1,1)\,\yng(1,1)\,\yng(1,1)}\quad
   O_8=O_{\tiny\yng(2,2,2,2,2,2,2,1,1,1,1,1,1,1),\,\yng(2,2,2,2,1,1,1,1,1,1,1,1,1)\,\yng(2)\,\yng(1,1)}
$$
$$
   O_9=O_{\tiny\yng(2,2,2,2,2,2,2,1,1,1,1,1,1,1),\,\yng(2,2,2,2,2,1,1,1,1,1,1,1)\,\yng(1,1)\,\yng(1,1)}\quad
O_{10}=O_{\tiny\yng(2,2,2,2,2,2,2,1,1,1,1,1,1,1),\,\yng(2,2,2,2,2,1,1,1,1,1,1,1)\,\yng(1,1)\,\yng(2)}\quad
O_{11}=O_{\tiny\yng(2,2,2,2,2,2,2,1,1,1,1,1,1,1),\,\yng(2,2,2,2,2,1,1,1,1,1,1,1)\,\yng(2)\,\yng(1,1)}\quad
O_{12}=O_{\tiny\yng(2,2,2,2,2,2,2,1,1,1,1,1,1,1),\,\yng(2,2,2,2,2,1,1,1,1,1,1,1)\,\yng(2)\,\yng(2)}
$$
$$
O_{13}=O_{\tiny\yng(2,2,2,2,2,2,2,1,1,1,1,1,1,1),\,\yng(2,2,2,2,2,1,1,1,1,1,1,1)\,\yng(1,1)\,\yng(1,1)}\quad
O_{14}=O_{\tiny\yng(2,2,2,2,2,2,2,1,1,1,1,1,1,1),\,\yng(2,2,2,1,1,1,1,1,1,1,1,1,1,1)\,\yng(1,1)\,\yng(1,1)}\quad
O_{15}=O_{\tiny\yng(2,2,2,2,2,2,2,1,1,1,1,1,1,1),\,\yng(2,2,2,2,1,1,1,1,1,1,1,1,1)\,\yng(1,1)\,\yng(1,1)}\quad
O_{16}=O_{\tiny\yng(2,2,2,2,2,2,2,1,1,1,1,1,1,1),\,\yng(2,2,2,2,1,1,1,1,1,1,1,1,1)\,\yng(1,1)\,\yng(2)}\, .
$$
When computing the dilatation operator, we assume that $b_1\ll b_0$, $b_0=O(N)$ and $b_1=O(N)$. The spectrum of the
dilatation operator that we obtain, when diagonalized numerically, does not reproduce the spectrum of a set of
decoupled oscillators. We do obtain a set of energy levels that is very well approximated by a linear spectrum
$E_n=\omega n$ with $\omega$ given by the average (over $n$) of $E_{n+1}-E_n$. However, $E_{n+1}-E_n$ is not
exactly equal to $8g_{YM}^2$ - it fluctuates around this value. We have also numerically verified that after 
invoking the approximations spelled out at the end of section 3.1, we do indeed obtain equation (\ref{recursion_j3}) 
and hence with these approximations the spectrum of the dilatation operator is again reproduced by a collection of 
decoupled oscillators. Thus, it is only after invoking the approximations of section 3.1 that we definitely obtain 
an integrable system.

The same conclusion is reached by studying the simpler system $m=2$, $p=1$, which involves 8 operators.

\end{document}